\documentclass[12pt]{article}
\usepackage{a4wide}
\usepackage{amssymb}

\def\bm#1{\mbox{\boldmath{$#1$}}}
\def\Mfunction#1{\mathop{\rm #1}\nolimits}

\def\A{{\cal A}}
\def\D{{\cal D}}
\def\H{{\cal H}}
\def\La{{\cal L}}

\def\O{{\cal O}}

\def\p{\partial}
\def\a{\alpha}
\def\b{\beta}
\def\d{\delta}

\def\f{\varphi}
\def\F{\Phi}

\def\l{\lambda}
\def\L{\Lambda}
\def\r{\rho}
\def\g{\gamma}

\def\o{\omega}

\def\s{\sigma}
\def\ra{\rightarrow}

\newcommand{\be}{\begin{equation}}
\newcommand{\ee}{\end{equation}}
\newcommand{\bea}{\begin{eqnarray}}
\newcommand{\eea}{\end{eqnarray}}

\begin{document}

\title{The relativistic virial theorem and scale invariance}

\author{Jos\'e Gaite
\\
\small\em
IDR, Universidad Polit\'ecnica de Madrid, E-28040 Madrid, Spain
}

\date{May 29, 2013}
\maketitle

\begin{abstract}
  The virial theorem is related to the dilatation properties of bound
  states. This is realized, in particular, by the Landau-Lifshitz formulation
  of the relativistic virial theorem, in terms of the trace of the
  energy-momentum tensor. We construct a Hamiltonian formulation of
  dilatations in which the relativistic virial theorem naturally arises as the
  condition of stability against dilatations. A bound state becomes scale
  invariant in the ultrarelativistic limit, in which its energy vanishes.
  However, for very relativistic bound states, scale invariance is broken by
  quantum effects and the virial theorem must include the energy-momentum
  tensor trace anomaly.  This quantum field theory virial theorem is directly
  related to the Callan-Symanzik equations. The virial theorem is applied to
  QED and then to QCD, focusing on the bag model of hadrons. In massless QCD,
  according to the virial theorem, $3/4$ of a hadron mass corresponds to
  quarks and gluons and $1/4$ to the trace anomaly.
\end{abstract}

\section{Introduction}

The classic virial theorem has been very useful in physics, in particular, in
astrophysics, to determine the equilibrium and stability of dynamical systems
\cite{Collins}. The theorem is especially useful for systems interacting
through potentials that are homogeneous functions of the interparticle
distances (e.g., power law potentials): then the theorem states a simple
relation between the long-time averages of the kinetic and potential energies
of a system, namely, $2 K = n U$, where $n$ is the homogeneity degree (e.g.,
the exponent of the power law).  Of course, the most relevant potentials to
consider are the quadratic potential of harmonic oscillations ($n=2$) and the
inverse distance law of the Newton or Coulomb potentials ($n=-1$). The virial
theorem for homogeneous potentials can be regarded as a consequence of
mechanical similarity, in other words, of scale invariance in mechanics
\cite{LL1}. Indeed, several authors have studied the relation of the virial
theorem to scale invariance \cite{Low,AB,Kleban} and its connection with
Noether's theorem \cite{Kampen,Nach,Blud-Ke}.

The classic virial theorem has been generalized in several ways
\cite{Collins}. The obvious generalizations pertain to relativistic mechanics
and quantum mechanics. Whereas the generalization of the virial theorem to
non-relativistic quantum mechanics
presents no special problems, the generalization to relativity is non-trivial,
because the concepts of force and potential are unsuitable for describing
relativistic interactions.  Notwithstanding, there are relativistic
formulations of the virial theorem.  A virial theorem in electrodynamics that
expresses the energy in terms of the energy-momentum tensor trace already
appears in the classic textbook by Landau and Lifshitz \cite{LL2}.  This
relativistic virial theorem is generalizable, in principle, to other
interactions \cite{Collins}.

The relativistic virial theorem has featured in several works and some papers
have been specifically devoted to it, both in classical field theory
\cite{Raf,Brack} and in quantum field theory \cite{DP}.  However, while the
relation of the classic virial theorem to scale invariance is well
established, the relation to scale invariance of the relativistic virial
theorem, especially, in the Landau-Lifshitz formulation, is hardly studied.
In this paper, we focus on this relation and adopt a fundamental standpoint.
Therefore, our approach to the relativistic virial theorem is closer to the
one of Ref.~\cite{Brack} and especially to the one of Ref.~\cite{DP} than to
the one of Ref.~\cite{Raf}.  Our main purpose is to establish the fundamental
role of scale transformations and scale invariance in the relativistic virial
theorem. For this purpose, we find it useful to proceed from the classic
theorem to the relativistic theorem and, finally, to the virial theorem in
quantum field theory (QFT).  In this process, the classic theorem $2 K = n U$,
which allows for any homogeneity degree $n$, must be restricted to
relativistic interactions mediated by a field that becomes in the
non-relativistic limit just an inverse-law potential, so the only allowed
value is $n=-1$.  In other words, the virial theorem is restricted to
interactions mediated by massless gauge fields, namely, the electromagnetic
interaction and the strong interaction described by quantum chromodynamics
(QCD). The latter has special features, as will be explained.

The classic virial theorem includes the gravitational interaction, but the
relativistic formulation of gravity necessarily leads to General Relativity,
which is a sort of interaction mediated by a massless gauge field, although
not of the ordinary type. In fact, there have been attempts to formulate the
virial theorem in General Relativity \cite[p~27]{Collins}, but they neglect
its relation to other gauge theories. Some of the results of the present paper
can be applied to General Relativity, but this theory has distinct features
and, in particular, in it the concept of gravitational energy as well as the
concept of scale invariance become very subtle and difficult to handle.  These
problems are beyond the scope of this paper and are left for future work.

We begin in Sec.~\ref{rel-em} with a Hamiltonian formulation of the virial
theorem that directly arises from the notion of time-averaged scale
invariance.  This formulation is applied to the electromagnetic interaction in
special relativity, considering, first, the action-at-a-distance interaction
of particles and, second, the full field theory, following Landau and Lifshitz
\cite{LL2}.  Next, in Sec.~\ref{HFT}, we construct a fully general derivation
of the field theory virial theorem that is based on scale invariance, by
generalizing the Hamiltonian formulation of the virial theorem for a system of
particles.  Some consequences of the virial theorem for bound states of
ultrarelativistic particles appear in Sec.~\ref{ultra}.  As these bound states
sustain considerable quantum effects, they call for relativistic quantum
mechanics.  The quantum field theory virial theorem is studied in
Sec.~\ref{non-inv}, applying it, first, to quantum electrodynamics (QED) and,
second, to quantum chromodynamics (QCD), in which the energy-momentum tensor
trace anomaly plays a crucial role.

\section{The relativistic virial theorem} 
\label{rel-em}

The virial theorem is regarded by Landau and Lifshitz \cite{LL1} as a
consequence of scaling in Lagrangian mechanics.  This line of thought has been
followed by several authors \cite{Low,AB,Kleban,Blud-Ke}, some of whom relate
the virial theorem to the Noether theorem.  In contrast to Lagrangian scaling,
we now introduce the general theory of dilatations in the Hamiltonian
formalism, which provides a concise and powerful formulation of the virial
theorem.

In Hamiltonian mechanics, canonical transformations are the most general
transformations that preserve the phase space structure \cite{LL1}. A
canonical transformation of the phase space $(q,p)$ is defined by its
infinitesimal generator, which is just any function $F(q,p)$.  For simplicity,
we use single variables $q$ and $p$, but each denotes the full set of
coordinates or momenta. The transformed phase space variables are:
\begin{eqnarray*}
Q = q + \d_Fq, \quad \d_Fq = \{q,F\} = \frac{\p F}{\p p}\,,\\ 
P = p + \d_Fp, \quad \d_Fp = \{p,F\} = -\frac{\p F}{\p q}\,, 
\end{eqnarray*}
where we have introduced Poisson brackets $\{\cdot,\cdot\}$. Naturally, the
phase-space volume element $dqdp$ is unchanged. 
Among canonical transformations, an important role is played by point
transformations, namely, canonical transformations induced by transformations
of the coordinates $q$ only. They are generated by $F= f(q)p$, where $f$ is
any function. If $F= q p$, we have
\begin{equation}
\d_Fq = q, \quad \d_Fp = -p,
\label{dFqp}
\end{equation}
that is, a homogeneous dilatation of $q$ and the corresponding homogeneous
contraction of $p$, which together preserve the phase space volume.  But,
normally, the dilatation of $q$ is not a symmetry, that is,
$$
\d_FH = \{H,F\} = q\frac{\p H}{\p q} - p\frac{\p H}{\p p} \neq 0,
$$
and $F$ is not a constant of the motion, that is, $\dot{F} = \{F,H\} \neq 0,$
unless the Hamiltonian $H$ is very special (e.g., $H=F$).  Nevertheless, if
both $q$ and $p$ are bounded, the temporal average of $\dot{F}$ vanishes in
the long run, so that, on the average,
\begin{equation}
\{F,H\} = p \dot{q} - q\frac{\p H}{\p q} = H + L - q\frac{\p H}{\p q} = 0.
\label{genvir}
\end{equation}
Note that the coordinates $q$ of a system of particles can only be bounded in
their rest frame, namely, the frame in which their total momentum is zero.

In many mechanical problems, $L(q,\dot{q}) = K(\dot{q}) - U(q)$, namely, there
is a separation between kinetic and potential energies, and $K$, in addition,
is quadratic in the velocities. Then, $H = K + U$, so
$$
H + L - q\frac{\p H}{\p q}  = 
2K - q\frac{\p U}{\p q} = 0.
$$
The term $q\,\p_q U$ is called virial.  Furthermore, when $U(q)$ is a
homogeneous functions of $q$ of degree $n$, Euler's theorem on homogeneous
functions implies that $q\,\p_q U=nU$, so there follows the standard virial
theorem, $2K=nU$ or, in terms of the total energy $E$, $K=nE/(n+2)$
\cite{Collins,LL1}.  This theorem is a consequence of the homogeneity of $K$
in $\dot{q}$ and the homogeneity of $U$ in $q$, which also imply mechanical
similarity: the equations of motion permit a set of geometrically similar
motions, such that the paths are geometrically similar and the times of motion
between corresponding points are in a constant ratio \cite{LL1}.  For example,
when $n=-1$, as in the Newton or Coulomb potentials, to $q \ra lq$ correspond
$t \ra l^{3/2} t$ (Kepler's third law) and $E \ra E/l$.

Unfortunately, the Lagrangian in relativistic mechanics is not of the form
$L(q,\dot{q}) = K(\dot{q}) - U(q)$ with $K$ homogeneous in $\dot{q}$.
Nevertheless, the classical virial theorem is easily extended to the case of a
relativistic particle under external forces \cite{Gold}.  Moreover, the
Hamiltonian virial theorem, Eq.~(\ref{genvir}), is very general and covers the
many-body problem in relativistic electrodynamics, as the prototype of
relativistic interactions.  Before considering this problem, let us consider
two pertinent extensions of the Hamiltonian virial theorem. The first one is
the extension to quantum mechanics. This extension is straightforward, because
the Hamiltonian formulation of canonical transformations is easily transferred
to quantum mechanics, just by transforming Poisson brackets to
commutators. However, one has to take care of operator ordering problems, for
example, by symmetrizing phase space functions with respect to $q$ and $p$.
Any phase space function that can be expanded in powers of $q$ and $p$ can be
symmetrized by symmetrizing every term of the expansion (which is a
polynomial). In particular, the dilatation generator becomes $F= (q p + pq)/2$
(naturally, $p= -i\hbar\,\p/\p q$, in the coordinate representation).

As a second extension of the virial theorem, this theorem can also be extended
to canonical transformations other than dilatations. Indeed, the long-time
average of $\dot{F}$ vanishes for any bounded function $F$ and so does the
average of $\{F,H\}$. Therefore, we have an infinite number of (average)
relations. However, general canonical transformations, with generators $F$
that depend on $p$ arbitrarily, have little physical significance. In
contrast, $F= f(q)p$,
namely, the set of point transformations, includes rotations and, furthermore,
arbitrary deformations of the geometrical ``shape'' of the mechanical system.
These transformations give rise to an extension of the standard virial
theorem, namely, the {\em tensor virial theorem} \cite{Collins},
\label{TVT} 
to be discussed further in page \pageref{TensorVT}.

\subsection{Virial theorem for the electromagnetic interaction}
\label{em-virial}

The $N$-particle electromagnetic Lagrangian is \cite{LL2,Jack}
$$
L = 
\sum_{a=1}^N \left(-m_a c^2 \sqrt{1-\frac{\bm{v}_a^2}{c^2}} +
\frac{e_a}{c} \bm{v}_a \cdot \bm{A}(\bm{x}_1,\ldots, \bm{x}_N)  - e_a
\F(\bm{x}_1,\ldots, \bm{x}_N) \right) ,
$$
where $\bm{v}_a = \dot{\bm{x}}_a\,.$ 
The corresponding Hamiltonian is 
\begin{equation}
H = \sum_{a=1}^N \left( \sqrt{(c\bm{p}_a-e_a\bm{A})^2+ m_a^2 c^4} + e_a \F \right) .
\label{Hem}
\end{equation}
To allow for interaction between the particles, the electromagnetic potentials
must fulfill the D'Alembert wave equation with a source given by the
electromagnetic current produced by the particles themselves, so the
potentials are functions of the particle trajectories. If the symmetric
Green's function (half-sum of advanced and retarded functions)
is used to solve the wave equation, the Lagrangian or
Hamiltonian correspond to {\em action-at-a-distance} electrodynamics
\cite{a-d,Barut}, in which there is no radiation and, therefore, no
need of the electromagnetic field Lagrangian.

Neither the $N$-particle Lagrangian nor the Hamiltonian can be separated into
kinetic and potential terms, but we can apply Eq.~(\ref{genvir}) nonetheless.
Given that
$$
\left(\frac{\p H}{\p \bm{x}_a}\right)_{\bm{p}} = -\dot{\bm{p}}_a =
-\left(\frac{\p L}{\p \bm{x}_a}\right)_{\bm{v}} ,
$$
Eq.~(\ref{genvir}) writes
\begin{eqnarray}
H + L +
\sum_a 
\bm{x}_a\cdot
\frac{\p }{\p \bm{x}_a}
\left(\sum_b \frac{e_b}{c} \bm{v}_b \cdot \bm{A} - e_b \F \right)
= \nonumber\\
H + 
\sum_a \left[-m_a c^2 \sqrt{1-\frac{\bm{v}_a^2}{c^2}} +
\left(1 + \bm{x}_a\cdot \frac{\p }{\p \bm{x}_a}\right)
\left(\sum_b \frac{e_b}{c} \bm{v}_b \cdot \bm{A} - e_b \F \right)
\right] = 0,
\label{EMvir}
\end{eqnarray}
where the derivatives are to be taken for constant $\bm{v}_b$.  Notice that
the action of the operator $\sum_a (1 + \bm{x}_a\cdot \p/\p \bm{x}_a)$ on a
homogeneous function of $\bm{x}_1,\ldots, \bm{x}_N$ of degree $n=-1$ gives a
null result, according to Euler's theorem. In other words, if we assume that
the vector and scalar potentials can be written as homogenous functions of the
coordinates with degree $n=-1$, each electromagnetic potential and its
corresponding virial cancel one another.  So we obtain the energy
\begin{equation}
E = \sum_a m_a c^2 \sqrt{1-\frac{\bm{v}_a^2}{c^2}}\,.
\label{E}
\end{equation}
This is the relativistic virial theorem for a system of particles in classical
electrodynamics.  If external forces were necessary to confine the particles
and, in particular, the forces consisted in a constant pressure $P$ exerted on
the system's surface, then $E$ should be replaced with $E - 3PV$, where $V$ is
the system's volume \cite[\S\hspace{2pt}35]{LL2}.

Although the electromagnetic parameters are absent from Eq.~(\ref{E}), their
effect is implicitly included: notice that $E < \sum_a m_a c^2$, as
corresponds to a bound state.  The bound state becomes non-relativistic for
low velocities, when $|E - \sum_a m_a c^2| \ll \sum_a m_a c^2$ and
Eq.~(\ref{E}) reduces to the classical virial theorem $E - \sum_a m_a c^2
\approx - \sum_a m_a v_a^2/2= -K $.
However, the relativistic dynamics has lost the similarity of the classical
dynamics under space dilatations and corresponding time dilatations.  This
similarity is lost even assuming that $\F$ and $\bm{A}$ are homogenous
functions with degree $n=-1$.  Moreover, in relativistic mechanics, there is
no similarity even for free particles ($e_a=0$), because of the form of the
Hamiltonian.  (Of course, similarity is recovered for low velocities, but the
energy only scales after subtracting the particles' rest energy $\sum_a m_a
c^2$.)
\label{no-sim}
However, there is a relativistic notion of mechanical similarity \cite{AB}, in
which space and time are equally dilatated, so velocities are unchanged.
Under this similarity, masses cannot be held constant and must be contracted
so that they transform like energies.  Therefore, the nature of the particles
is changed and the similarity does not relate different motions of the same
system.  Nevertheless, this similarity leads to Eq.~(\ref{E}) and, in
addition, it also leads to the classic virial theorem in the non-relativistic
limit \cite{AB}.

The equations of motion of action-at-a-distance electrodynamics are not
ordinary differential equations but differential-difference equations, and it
is hard to find solutions of them.  A simple solution of the relativistic
two-body problem is provided by two opposite charges in circular motion, with
calculable radii and angular velocity \cite[p.~223]{a-d}.  This solution
fulfills Eq.~(\ref{E}) (no temporal average is needed).  At any rate, it is
natural to consider the electromagnetic field's dynamics and the radiation
effects.  However, given that these effects appear, in an expansion in powers
of $v/c$, only in the third order, it is possible to describe the
electromagnetic interaction of particles with standard ordinary differential
equations based on potentials of order $(v/c)^2$ \cite{LL2,Jack,a-d}.  These
potentials, added to a relativistic kinetic term expanded to the same order,
give rise to the well-known Darwin Hamiltonian and Lagrangian.  In this
approximation, both the vector and scalar potentials are homogeneous functions
of degree $n=-1$, so the virial theorem (\ref{E}), expanded to terms of second
order, holds.

Regarding quantum mechanics, the relativistic virial theorem holds, by
canonical quantization of Hamiltonian systems, as said before.  Indeed,
quantum relativistic versions of the virial theorem have appeared in the
literature \cite{Aus,Kor}. However, they are meant to be applied in nuclear
physics and they only consider a simple two-quark problem with a
phenomenological scalar potential $U$ (notice that the fundamental theory of
strong interactions, namely, QCD, includes a {\em vector} potential, like
QED).  The scalar potential is the ``Cornell potential,'' which is the sum of
a Coulomb term and a confining linear term.  Lucha and Sch\"oberl \cite{Aus},
in particular, obtain
\begin{equation}
\langle \bm{x}\cdot\nabla U(\bm{x}) \rangle =
c \langle \frac{\bm{p}^2}{\sqrt{\bm{p}^2+(m_1c)^2}} +
\frac{\bm{p}^2}{\sqrt{\bm{p}^2+(m_2c)^2}} 
\rangle, 
\label{centrifugal}
\end{equation}
and, hence, 
\begin{equation}
E = \langle \bm{x}\cdot\nabla U(\bm{x}) \rangle +
\langle U(\bm{x}) \rangle +
c^3 \langle \frac{m_1^2}{\sqrt{\bm{p}^2+(m_1c)^2}} +
\frac{m_2^2}{\sqrt{\bm{p}^2+(m_2c)^2}} 
\rangle, 
\label{L-S}
\end{equation}
where the expectations values are understood to be taken with respect to
normalized eigenstates.  Eq.~(\ref{centrifugal}) states the equality of the
expectations values of the centripetal and centrifugal virials.
On the other hand, on account of the relativistic identity
$$
1-\frac{\bm{v}^2}{c^2} = \frac{(m c)^2}{\bm{p}^2+(m c)^2}\,,
$$
Eq.~(\ref{L-S}) is a quantum version of Eq.~(\ref{EMvir}), such that the
temporal expectation values implicit in Eq.~(\ref{EMvir}) are replaced by
expectation values on stationary states and such that the system is restricted
to two particles with $\bm{A}=0$. In Eq.~(\ref{L-S}), the Coulomb part of the
potential and its virial indeed cancel one another, like in Eq.~(\ref{EMvir}).
At any rate, the strong interaction Hamiltonian employed by
Refs.~\cite{Aus,Kor} is not fully relativistic and only has relativistic
kinematics (the potential $U$ can be interpreted as the lowest order
slow-motion approximation of a relativistic interaction).
The formulation of a fully relativistic virial theorem for quark bound states
requires a QFT framework and is presented in Sec.~\ref{non-inv}.

Returning to classical electrodynamics, let us consider the full particle plus
field dynamics and its local conservation laws, namely, the local conservation
of energy and momentum.  Landau and Lifshitz's virial theorem \cite{LL2}
relies on this conservation law, expressed in terms of the energy-momentum
tensor.  This conservation law implies that the long-time average of the
stress tensor (the spatial part of the energy-momentum tensor) is
divergenceless; that is,
\begin{equation}
\p_j\overline{T_i^j}=0,
\label{pTij}
\end{equation}
where Latin indices denote spatial coordinates and the bar denotes the
long-time average.  Multiplying Eq.~(\ref{pTij}) by $x^i$ and integrating over
all space, one obtains (under suitable asymptotic conditions) that the space
integral of the stress-tensor trace average vanishes:
\begin{equation}
\int \overline{T_i^i}\, dV = 0.
\label{Tii}
\end{equation}
Actually, the vanishing of the integral of the {\em full} stress tensor for a
closed (self-contained) and {\em static} system was proved by Laue \cite{Laue}
at the beginning of relativity theory.%
\footnote{Laue's paper studies the energy and momentum of a closed static
  system and, according to Ohanian \cite{Ohanian}, it contains the {\em first}
  real proof of the mass-energy equivalence, $E = mc^2$.}
The vanishing of the integral of the time average of the stress tensor
constitutes the {\em tensor virial theorem}, which holds in classical
mechanics as well as in relativistic mechanics (Ref.~\cite{Collins},
\S\hspace{2pt}II.1 and \S\hspace{2pt}II.3). \label{TensorVT} It can be proved
by multiplying Eq.~(\ref{pTij}) by $x^k$, that is, using a generic index $k$
instead of the index $i$ used above, and integrating over all space.
Regarding the connection of the tensor virial theorem with arbitrary
transformations of spatial coordinates, already mentioned in page
\pageref{TVT}, this theorem can be understood as just a condition of {\em
  dynamical} equilibrium, namely, of stability of the averaged system shape
against deformations, as explained in Sec.~\ref{HFT}.
Of course, the tensor virial theorem implies the scalar
theorem (the vanishing of the trace), which can be understood as a condition
of stability of the system against dilatations.

For a system of $N$ bodies in electromagnetic interaction,
\begin{equation}
T_i^i = \sum_{a=1}^N \frac{m_a  {\bm{v}_a^2}}{\sqrt{1-{\bm{v}_a^2}/c^2}}\,
\d(\bm{x}-\bm{x}_a) + \frac{1}{2}(\bm{E}^2 + c^2\bm{B}^2).
\label{Tii-N}
\end{equation}
Therefore, the space integral of $T_i^i$ 
seems to be strictly positive and the virial theorem,
Eq.~(\ref{Tii}), cannot be fulfilled.
To interpret Eq.~(\ref{Tii-N}) and the corresponding virial theorem, 
let us make the orthogonal decomposition $\bm{E} = \bm{E}_\mathrm{L} +
\bm{E}_\mathrm{T}$, 
where $\bm{E}_\mathrm{L}= -\nabla \Phi$ and $\nabla \cdot \bm{E}_\mathrm{T}= 0$.
Then, the electromagnetic energy writes
\begin{equation}
  \frac{1}{2}\int dV (\bm{E}^2 + c^2\bm{B}^2) = \frac{1}{2}\int dV
  (\bm{E}_\mathrm{L}^2 +  \bm{E}_\mathrm{T}^2 + c^2\bm{B}^2),
\label{E_L-T}
\end{equation}
where
\begin{equation}
\frac{1}{2}\int dV \bm{E}_\mathrm{L}^2 = \frac{1}{8\pi}
\int dV dV'\,
\frac{\r(\bm{x})\r(\bm{x}')}{\left|\bm{x}-\bm{x}'\right|}
\label{Coul}
\end{equation}
is the Coulomb electrostatic energy.  While this electrostatic energy is
obviously positive for regular distributions, the electrostatic energy of a
system of $N$ positive and negative {\em point-like} charges can become
negative after the subtraction of the (infinite) self-energy of the charges,
which amounts to a renormalization of their masses \cite{LL2,Jack}.  On the
other hand, regarding the contribution of $\bm{E}_\mathrm{T}$ and $\bm{B}$ to
the energy, one must impose the absence of incoming or outgoing radiation
fields, which will produce contributions to $\bm{E}_\mathrm{T}^2 +
c^2\bm{B}^2$ of arbitrary magnitude.
Indeed, the proof of Eq.~(\ref{Tii}) requires the vanishing of certain surface
integral related to radiation fields \cite{LL2}.  As already said, the absence
of radiation is implicit in action-at-a-distance electrodynamics.  These
problems have already been noticed by Dudas and Pirjol \cite{DP}, who
emphasize the role of the Wheeler-Feynman time-symmetric formulation of
electrodynamics in their solution.

The four-dimensional trace is $T_\mu^\mu = -T^{00} + T_i^i$ [our metric's
signature convention is $(-,+,+,+)$].  Therefore, a condition equivalent to
(\ref{Tii}) that introduces the total energy is \cite{LL2}
\begin{equation}
E = -\int T_\mu^\mu \,dV
\label{E=T}
\end{equation}
(like before, the long-time average is implicit when not explicitly shown).
For a system of electromagnetically interacting particles, the electromagnetic
energy-momentum tensor is traceless and disappears from Eq.~(\ref{E=T}). This
absence of electromagnetic parameters at virial equilibrium is analogous to
the cancellation of the electromagnetic potentials with their corresponding
virials in Eq.~(\ref{EMvir}).  Indeed, Eq.~(\ref{E=T}) leads again to
Eq.~(\ref{E}) \cite{LL2}.

Notice that Eq.~(\ref{E=T}) makes no reference to particles, unlike
Eq.~(\ref{E}).  Therefore, it can be applied to fields forming a {\em single}
particle and then it plays a role in the famous problem of modeling the
electron or any charged particle in classical electrodynamics
\cite[Ch.~16]{Jack}. Indeed, Eq.~(\ref{E=T}) connects the particle's energy
with the trace of the energy-momentum tensor of the Poincar\'e stresses, since
the electromagnetic energy-momentum tensor is traceless.  If we assume, for
simplicity, that the energy-momentum tensor of the Poincar\'e stresses has no
traceless part, namely, that it is proportional to $g_{\mu\nu}$, then the
total energy-momentum tensor fulfills $T^{\mu\nu}=T^{\mu\nu}_{\mathrm{em}} +
T_\a^\a g^{\mu\nu}/4$, where $T^{\mu\nu}_{\mathrm{em}}$ is the energy-momentum
tensor of the electromagnetic field. Therefore, in the rest frame,
\begin{equation}
E = \int \left(T^{00}_{\mathrm{em}} +  \frac{g^{00}}{4}T_\a^\a\right) dV =
\int T^{00}_{\mathrm{em}} \,dV + \frac{E}{4}\,,
\label{virialE}
\end{equation}
so three fourths of the particle's rest energy come from its electromagnetic
energy and the remaining one fourth from the Poincar\'e stresses ($3/4$ is
actually the ratio of electromagnetic mass to electromagnetic inertia that
constitutes the infamous $4/3$ problem, solved by the introduction of the
Poincar\'e stresses \cite[Ch.~16]{Jack}).  If the energy-momentum tensor
corresponding to the Poincar\'e stresses, $T^{\mu\nu}_{\mathrm{P}}$, has a
non-vanishing traceless part, this part is on the same footing as the
traceless energy-momentum tensor of the electromagnetic field, so both
together make up the three fourths of the total particle's rest energy.  The
contribution to the energy of the traceless part of the energy-momentum tensor
for the Poincar\'e stresses is the space integral of
\begin{equation}
T^{00}_{\mathrm{P}} - T_{\mathrm{P}}{}^\a_\a \,\frac{g^{00}}{4} =  
\frac{3}{4}T^{00}_{\mathrm{P}} + \frac{1}{4} T_{\mathrm{P}}{}^i_i \,.
\label{nullE}
\end{equation}
This integral is non-negative, despite that $\int T_{\mathrm{P}}{}^i_i\,
dV<0$, as is necessary to have cohesive Poincar\'e stresses, that is to say,
as is necessary for the total stress tensor to fulfill Eq.~(\ref{Tii}).  The
non-negativity of the right-hand side of Eq.~(\ref{nullE}) can be proved by
invoking the {\em null energy condition} (deduced by continuity from the
non-negativity of the energy density
\cite[p.\ 89]{Hawk-Ell}).  In conclusion, the proportion of energy due to the
Poincar\'e stresses is at least one fourth, and the proportion of
electromagnetic energy can be equal or smaller than three fourths.

In a single-particle model, the charge distribution is continuous and can be
attributed to a charged field or fluid.  An interesting classical electron
model is Bialynicki-Birula's \cite{BB}, in which the electron consists of a
perfect, charged fluid, with energy density $\r$ and pressure $P$, and the
electromagnetic field. The traceless part of the fluid energy-momentum tensor
is proportional to $\r + P$ and contributes to the energy with $(3/4)\int (\r
+ P) dV$, which is non-negative, despite that $P<0$ everywhere.

Notice that a particle model with a continuous charge distribution makes as
much sense for a composite particle as for an elementary particle, since there
is no quantization of charge in classical electrodynamics.  For consistency
with electron modeling in QED, let us assume that matter is described by a
Dirac field $\psi$ with the standard Lagrangian.  Then,
$$
T_{\mu\nu}= \frac{i}{2} \,c \left(
\bar{\psi} \g_{(\mu}\D_{\nu)} \psi -\D^*_{(\mu}\bar{\psi} \g_{\nu)} \psi 
\right),
$$
where $\D_{\mu} = \hbar \p_{\mu} + i(e/c)A_{\mu}\,,$ and $\D^*$ is its complex
conjugate.  Therefore,
\begin{equation}
T_{\mu}^{\mu}= \frac{i}{2} \,c \left(
\bar{\psi} \g^{\mu}\D_{\mu} \psi - \D^*_{\mu}\bar{\psi} \g^{\mu} \psi
\right) = 
-m c^2\bar{\psi}\psi 
\label{T-fermi}
\end{equation}
and, according to Eq.~(\ref{E=T}), the energy of a composite or elementary
particle is
\begin{equation}
E = m c^2 \int \bar{\psi}\psi  \,dV.
\label{fermi-vir}
\end{equation}
This equation is related to Fock's old result for the Dirac equation in an
external, central Coulomb field \cite{Fock} and also to the more general virial
theorem of Rose and Welton \cite{RW} (see Ref.~\cite{Brack} as well).

Since we are considering the Dirac field $\psi$ as a {\em classical} field,
constituting a sort of matter fluid, $n(\bm{x})=\bar{\psi}\psi$ is the total
particle number density, with equal weight for particles and antiparticles,
and computed in the local reference frame.
Equation (\ref{fermi-vir}) might seem to imply that the bound-state energy is
just the number of particles times the rest energy per particle, 
as if they were free and at rest,
but it does not, because $n(\bm{x})$ must be computed in the {\em local}
reference frame. When $n(\bm{x})$ is referred to the laboratory frame, namely,
the bound-state rest frame, then
$$mn(\bm{x}) \ra r(\bm{x})\sqrt{1-v(\bm{x})^2/c^2},$$ 
where $r(\bm{x})$ is the ordinary non-relativistic mass density and
$v(\bm{x})$ is the velocity of the matter-fluid element $dm = r(\bm{x}) \,dV$.
Thus, we obtain
\begin{equation}
E 
= c^2 \int \sqrt{1-\frac{v(\bm{x})^2}{c^2}}\,r(\bm{x}) \,dV,
\label{E-int}
\end{equation}
which is a continuous form of Eq.~(\ref{E}).
Naturally, this virial theorem applies only to bound-state solutions of the
nonlinear equations of the classical electrodynamics of the Dirac field, which
are hardly explored (see, for example, Ref.\ \cite{CED}).  As these equations
have stable bound-state solutions, the corresponding single particle models do
not need extraneous Poincar\'e stresses.

\subsection{Hamiltonian field theory formulation}
\label{HFT}

We can obtain a general field-theory virial theorem within the Hamiltonian
formalism by generalizing the derivation of the virial theorem for a finite
number of degrees of freedom in Sect.~\ref{rel-em}. Let us consider a generic
field, which we denote $\f$ but which can comprise a set of independent fields
(it can be a vector field, etc), and consider its Lagrangian and Hamiltonian
densities, respectively, $\La$ and $\H$.  The associated field momentum
density is
$$
\pi = \frac{\p \La}{\p \dot\f}\, .
$$
In particle mechanics, dilatations simply act on the coordinates as $q \ra lq$
and on the momenta as $p \ra p/l$, which in infinitesimal form are given by
Eq.~(\ref{dFqp}). In field theory, dilatations primarily act on space
coordinates and, through them, on field coordinates $\f$ and momenta $\pi$.
Therefore, the infinitesimal generator of the finite dilatation $\f(\bm{x})
\ra l^D \f(l\bm{x})$, where $D$ is the dimension matrix, is given by the Lie
derivative
$$
\d\f = D \f - x^i \p_i \f, 
$$ 
The dimension matrix $D$ can be assumed to be diagonal, that is to say, $\f$
can be assumed to be formed by eigenstates of $D$.  The Lie derivative $\d\f$
differs from $\d q$ in Eq.~(\ref{dFqp}) in the presence of the transport term,
with $x^i \p_i\,,$ and also in the presence of $D$, which generalizes the
trivial dimension of $q$.  The
canonical generator of dilatations $F$ such that $\d\f = \d F/\d\pi$ is
\begin{equation}
F = \int \pi (D \f - x^i \p_i \f) \, dV. 
\label{F-field}
\end{equation}
Therefore,
$$
\d\pi = -\frac{\d F}{\d\f} = (-3-D) \pi - x^i \p_i \pi\,,
$$
up to the vanishing of a surface integral. We deduce that both $\d\f$ and
$\d\pi$ constitute the infinitesimal generators of Hamiltonian
dilatations. Note that, under a finite dilatation, $\pi(\bm{x}) \f(\bm{x}) \ra
l^{-3} \pi(l\bm{x}) \f(l\bm{x})$, so that $F$ is invariant.

The generating function $F$ given by Eq.~(\ref{F-field}) is the integral of
the sum of two parts, one corresponding to the intrinsic change of the field
and another corresponding to the spatial transport. The latter can be written
as $x^i {{T_\mathrm{C}}^0}_i,$ where ${{T_\mathrm{C}}^{\mu}}_\nu$ comprises
the four conserved currents associated to space-time translations by Noether's
theorem \cite{W}; that is to say, it is the {\em canonical} energy-momentum
tensor.  For example, for a scalar field $\f$ with $\La= -\p^\mu \f \p_\mu
\f/2 - {\cal V}(\f)$,
$${{T_\mathrm{C}}^{\mu}}_\nu= \p^\mu \f \p_\nu \f + \d^{\mu}_{\nu} \La,$$ 
so 
$${{T_\mathrm{C}}^0}_{i}= -\dot{\f} \p_i \f = -\pi \p_i \f\,.$$
For the electromagnetic field,
$${{T_\mathrm{C}}^{\mu}}_\nu= F^{\mu\r}\p_\nu A_\r+ \d^{\mu}_{\nu} \La\,,$$ 
so, in the Hamiltonian gauge $A_0=0$,
$${{T_\mathrm{C}}^0}_{i} = -\dot{A}_j \p_i A^j = -\pi_j \p_i A^j.$$
Therefore, in general,
\begin{equation}
F = \int (\pi D \f + x^i {{T_\mathrm{C}}^0}_i) \, dV. 
\label{F-field_1}
\end{equation}
However, ${{T_\mathrm{C}}^0}_i$ can be redefined by adding to it $\p_j t^j_i$,
where $t^j_i$ is an arbitrary function of $\f$ and $\pi$.  With the
appropriate choice of $t^j_i$, one can cancel the first summand in the
right-hand side of Eq.~(\ref{F-field_1}) (up to a surface integral).
In other words, there is always an ``improvement'' of the energy-momentum
tensor such that the dilatation generator becomes
$$
F = \int x^i {T^0}_i\, dV.
$$ 
This connects the Hamiltonian formulation of the virial theorem with Landau
and Lifshitz's proof \cite{LL2}. Indeed, using the conservation law
$\p_{\mu}T^{\mu\nu}=0,$
$$
\dot{F} = \int {T^i}_i\, dV.
$$
The vanishing of the temporal average of $\dot{F}$ and, therefore, of the
spatial integral of the temporal average of ${T^i}_i$
give rise to the virial theorem, Eq.~(\ref{Tii}).  

Notice that one can also consider, instead of dilatations, general
(anisotropic) coordinate transformations and so derive the tensor virial
theorem.

The general condition of exact scale invariance is $\d_FH = -\dot{F} =0$, or
\begin{equation}
\dot{F}= \int {T^i}_i\, dV = 0,
\label{F-Tii}
\end{equation}
without averaging. This condition is not fulfilled by general field
configurations of normal field theories.  Naturally, $\dot{F}$ must vanish for
any static field configuration and hence so does the integral of the stress
tensor trace.  This is in accord with Laue's theorem (Sec.~\ref{em-virial}),
applicable to any static relativistic system and, in particular, to any model
of an elementary particle, such as the electron.  For example, for
Bialynicki-Birula's electron model, the condition (\ref{F-Tii}) is indeed
fundamental for relativistic invariance \cite{BB}.

Let us remark on one interesting consequence of Eq.~(\ref{F-Tii}) for static
field configurations.  In the case of a scalar field with $\La= -\p^\mu \f
\p_\mu \f/2 - {\cal V}(\f)$, a static field has
\begin{equation}
{{T_\mathrm{C}}^i}_i = \frac{2-d}{2} (\nabla\f)^2 - d \,{\cal V}(\f),
\label{Tii-scalar}
\end{equation}
where $d$ is the space dimension (we simply use the canonical, unimproved
energy-momentum tensor, because the space integral of the stress tensor is not
altered by the improvement).  
With full generality, 
we can take ${\cal V}(\f) \geq 0$ and vanishing at its
absolute minima; namely, we assume that there are several absolute minima
with ${\cal V} = 0$. 
Then, Eqs.~(\ref{F-Tii}) and (\ref{Tii-scalar}) imply, for $d \geq 2$, that
$\f(\bm{x})$ is constant and equal to its value at one of the minima.  The
absence of localized static solutions
of scalar field theories in $d \geq 2$ is known as the Hobart-Derrick theorem
\cite{Hobart,Derrick}
and is usually proved by direct scaling of $\f(\bm{x})$.  The generalization
of this theorem to other field theories more complicated than the scalar field
theory is also given by Eq.~(\ref{F-Tii}), although it can be proved by direct
scaling of the appropriate field(s), case by case \cite[Ch~6]{Coleman}.

In the current treatment of scale transformations, we have chosen examples of
relativistic fields, but let us notice that there is no need to impose Lorentz
invariance to obtain the virial theorem.  When there is Lorentz invariance,
the Lagrangian formulation of scale transformations, in terms of $\La$, is
usually employed instead of the Hamiltonian formulation, because it is
covariant and, hence, explicitly relativistic.  The Lagrangian formulation of
scale transformations is based on {\em space-time} dilatations, such that
$$\d\f = D \f - x^\a \p_\a \f\,.$$ 
They coincide with space-only dilatations for static fields.  The {\em local}
current associated to space-time dilatations by Noether's theorem can always
be expressed as $j_\mathrm{D}^\mu = x_\nu T^{\mu\nu}$ and scale invariance can
be expressed in the local form $\p_{\mu} j_\mathrm{D}^\mu = 0$
\cite{Coleman,CCJ}.  Therefore, in field theory, scale invariance is generally
connected with the tracelessness of the energy-momentum tensor.
Although the conserved symmetric energy-momentum tensor of a scale-invariant
field theory is not necessarily traceless, it is always possible to
``improve'' it and convert it into one that is traceless, in addition to
symmetric and conserved \cite{Coleman,CCJ}. Then, the tracelessness of the
energy-momentum tensor implies, beyond Poincar\'e and scale invariance,
full invariance under the conformal group, which is obtained by adjoining
the discrete inversion to Poincar\'e and scale transformations. 
The sourceless Maxwell equations are, of course, conformal invariant and, in
this case, the energy-momentum tensor that results from symmetrizing its
canonical form is already traceless \cite{LL2}, so it needs no improvement.
However, such improvement is necessary in other field theories, e.g., in the
massless scalar field theory.

\subsection{Scale invariance in the ultrarelativistic domain}
\label{ultra}

Regarding the virial theorem for an $N$-particle bound state in
electrodynamics, Eq.~(\ref{E}), there arises the possibility of bound states
of vanishing energy, namely, states such that $E \ll \sum_a m_a c^2,$ as the
bound particles become ultrarelativistic and approach the speed of light.  In
one such bound state, the kinetic energy tends to infinity, but this is
compensated by a potential energy that tends to minus infinity, as the
particles approach one another and the system collapses.

The vanishing of the energy of an ultrarelativistic bound state is a
consequence of scale invariance.  As remarked in page \pageref{no-sim}, the
relativistic form of mechanical similarity involves the scaling of mass, since
mass is inextricably linked to energy in relativistic mechanics. Hence, full
similarity demands the absence of masses.  In the ultrarelativistic limit,
$p_a \gg m_ac$, the electrodynamics Hamiltonian (\ref{Hem}) becomes
$$
H = \sum_{a} |c\bm{p}_a-e\bm{A}| + e \F ,
$$ 
which is also obtained just by setting $m_a=0$.  The absence of masses
suggests scale invariance. Indeed, provided that $\bm{A}$ and $\F$ are
homogeneous functions of degree $n=-1$, $H$ transforms into $H/l$ under the
phase-space coordinate scaling $\bm{x} \ra l\bm{x}$ and $\bm{p} \ra \bm{p}/l$.
In consequence, a system of ultrarelativistic particles is such that the
energy is proportional to $p$ and, therefore, in the limit $p \ra \infty$,
a bound state must
have $E=0$ exactly and be scale invariant.
Otherwise, $E$ is a non-vanishing function of the $m_a$ and the bound state
is 
not scale invariant.

One scale-invariant relativistic state of vanishing energy is the vacuum,
which is neutral and has no measurable property. In fact, one can
imagine that a neutral $N$-particle system loses energy by radiation
and traverses a sequence of ultrarelativistic bound states of decreasing
energy that ends in the vacuum. However, the study of the last stages of
this process demands a quantum field theory treatment. As an example of
decay to vacuum, one can consider the annihilation of  positronium, but
positronium is a weakly coupled system, such that its annihilation takes
place before it enters the ultrarelativistic domain. Besides, one can
also consider ultrarelativistic states with non-zero charge. A simple
example of ultrarelativistic dynamics in atomic physics is briefly
studied below. This example is useful, in particular, to introduce the
phenomenon of {\em vacuum decay}, important in QCD (Sec.~\ref{QCD}).

Although, apparently, there are no electromagnetically bound particles in
ordinary matter that are fully relativistic, the fastest electrons of certain
atoms actually are ultrarelativistic.  Let us focus on one of the innermost
and fastest electrons of a heavy atom, and, for simplicity, consider it as if
it were not influenced by the other electrons; namely, let us consider the
one-electron Hamiltonian
\begin{equation}
H = c \sqrt{\bm{p}^2 + (mc)^2} - \frac{Ze^2}{4\pi r}\,
\label{E-1e}
\end{equation}
(the nucleus can be taken at rest and, hence, can be neglected). 
For a circular orbit, the radial equation of motion just states the equality
of centrifugal and Coulomb forces, which can be written in terms of the
respective virials as 
\begin{equation}
\frac{c\bm{p}^2}{\sqrt{\bm{p}^2 + (mc)^2}} =  \frac{Ze^2}{4\pi r}\,.
\label{force-eq}
\end{equation}
For a general orbit, the equality of virials holds only on a temporal average.
Given that
$$\frac{c\bm{p}^2}{\sqrt{\bm{p}^2 + (mc)^2}} = \bm{p}\cdot\bm{v} = 
\frac{m\bm{v}^2}{\sqrt{1-\bm{v}^2/c^2}}\,,$$ 
the temporal average of Eq.~(\ref{force-eq}) is a particularly simple case of
the virial theorem constituted by Eqs.~(\ref{Tii}), (\ref{Tii-N}),
(\ref{E_L-T}), and (\ref{Coul}), with $\bm{E}_\mathrm{T}= \bm{B} = 0$. Also,
note the connection with Eq.~(\ref{centrifugal}).  As regards the energy, the
virial theorem says that the ratio $E/(mc^2)$ equals the temporal average of
$\sqrt{1-\bm{v}^2/c^2} = mc/\sqrt{\bm{p}^2 + (mc)^2}$.

As usual, virial relations can be used to draw conclusions on the dynamics 
without solving the equations of motion.
In particular, from Eq.~(\ref{force-eq}) and 
$$
\frac{1}{\sqrt{\bm{p}^2 + (mc)^2}} \leq 
\frac{1}{p}\,,
$$
we deduce that
$c p \geq Ze^2/(4\pi r)$, on the average. The equality $c p= Ze^2/(4\pi r)$
takes place in the ultrarelativistic limit, in which $r \ra 0$, $p \ra
\infty$, $E \ra 0$, and scale invariance is approached.  For a circular orbit,
the angular momentum is $M=pr$, so we have the condition that $M \geq
Ze^2/(4\pi c)$.  Actually, when $M < Ze^2/(4\pi c)$, no orbit is stable and
the electron must fall in towards the nucleus, in a spiral trajectory
\cite[\S\hspace{2pt}39]{LL2}.  
In particular, as the electron approaches the nucleus and becomes
ultrarelativistic, its trajectory tends 
to a logarithmic spiral, which is {\em self-similar}.
Nevertheless, there are stable orbits for every $E > 0$, although, as $E \ra
0$, the only stable orbits are the circular orbits and they become just
marginally stable. For $E < 0$, there are no stable orbits.

Let us now consider the one-electron heavy atom in quantum mechanics, where
the uncertainty principle sets a lower limit to the atom's size; namely, both
$r$ and $E$ are bounded below and their lowest values correspond to the ground
state of the Hamiltonian.
This happens whether the electron is relativistic or not, but in the
relativistic case, namely, for the $H$ of Eq.~(\ref{E-1e}), a new feature
arises: low $M$ states and hence states with the lowest positive energies can
be unstable; that is to say, states with $M \sim \hbar$ are stable only if
$Z\a < 1$, where $\a = e^2/(4\pi\hbar c)= 1/137$.  This is confirmed by
solving the problem with a relativistic wave equation, for example, the
Klein-Gordon, Dirac
or Salpeter equations: the ground state is unstable for large $Z\a$, and the
stability bound to $Z\a$, which depends on the wave equation considered, is
always of the order of unity.%
\footnote{ The spectrum of the $H$ of Eq.~(\ref{E-1e}) has been studied by
  Herbst \cite{Herb}.  He proves that Eq.~(\ref{force-eq}) holds as an
  equation for the expectation values in eigenstates of $H$ and proves that
  the spectrum is non-negative for $Z\a \leq 2/\pi$, whereas it is unbounded
  below for $Z\a > 2/\pi$.}  This instability can be interpreted as a quantum
mechanical collapse in which the standard vacuum decays and gives rise to a
new, negatively charged vacuum \cite[\S\hspace{2pt}7]{Greiner}.  The charged
vacuum can be pictured as an electron cloud attached to the nucleus. This
cloud corresponds to the classical spiral trajectories that fall into the
nucleus and is also asymptotically self-similar.  

Let us remark that this vacuum instability occurs due to the presence of a
non-electromagnetic interaction, namely, the strong interaction, which 
keeps the positive charge $Ze$ within a nucleus sufficiently small
as to produce a very strong electromagnetic field.
In contrast, the electron-positron system does not have negative energy
states, owing to the smallness of $\a$. A proper study of this problem demands
quantum field theory methods, but the problem can be reduced, in a certain
approximation, to an equation similar to the Schr\"odinger equation for the
$H$ defined by Eq.~(\ref{E-1e}) \cite{FHM}. If there were electron-positron
negative energy states, the standard QED vacuum would not be stable against
condensation of electron-positron pairs. This vacuum decay does not occur in
QED, but quark-antiquark condensation and vacuum decay take place in QCD
\cite{FHM}.  This property of the QCD vacuum is crucial for hadron physics, as
studied in Sec.~\ref{QCD}.

Regarding scale invariance, the fundamental effect of quantum mechanics is to
introduce the new constant $\hbar$, which together with $c$ leaves only one
physical dimension, say, length.  Therefore, particle masses can be associated
with length scales, namely, their associated Compton wavelengths.  Let us
remark that this association is consistent with the Andersen-Baeyer \cite{AB}
relativistic similarity.  The Compton wavelength $\hbar/(mc)$ marks the scale
where the momentum or energy uncertainties are large enough to allow the
creation of a particle-antiparticle pair, so the very concept of particle
loses meaning. The relevant wavelength for an atom, namely, $\hbar/(mv) \sim
\hbar/(mcZ\a)$ is definitely larger than the electron's Compton wavelength
if $Z\a \ll 1$.  In the opposite case, $Z\a \gtrsim 1$, the electron becomes
ultrarelativistic and the potential of the (point-like) nucleus is strong
enough to induce the creation of electron-positron pairs.  As scale invariance
can only take place in the ultrarelativistic domain, it takes place for length
scales much smaller than $\hbar/(mcZ\a)$, where the one-electron description
is inadequate.  In general, any mass scale breaks scale invariance in
relativistic quantum mechanics, and scale invariance can only be recovered in
the ultrarelativistic domain, but in this domain the quantum effects
associated with particle creation take over.  It turns out that, in addition
to the explicit breaking of scale invariance by any mass, scale invariance is
{\em always} broken by quantum effects on small scales, even in massless
systems \cite{Coleman,CCJ}.  The failure of a classical symmetry due to
quantum effects is called a {\em quantum anomaly}.  The scale invariance
anomaly is important for bound states, especially in QCD, and features in the
formulation of a QFT virial theorem (see next section).

\section{The quantum field theory virial theorem} 
\label{non-inv}

In relativistic quantum mechanics, scale invariance is broken on scales of the
order of the particles' Compton wavelengths, as said above.  On the other
hand, on these scales, a bound system cannot be described in terms of a
definite set of particles that interact through a field, because the
Schr\"odinger equation for that set of particles neglects the possible
creation of more particles.  As is well known, relativistic quantum mechanics
leads to quantum field theory, in which particles and fields are on the same
footing, so the Schr\"odinger equation is best expressed in terms of all the
fields present.  Therefore, the virial theorem for a definite set of
elementary particles given by Eq.~(\ref{E}) is naturally replaced by the field
theory formulations given by Eq.~(\ref{Tii}) or Eq.~(\ref{E=T}).  The quantum
versions of these forms of the virial theorem can be derived by analogy with
the Landau-Lifshitz proof \cite{DP} or directly from Eq.~(\ref{F-Tii}),
yielding:
\begin{equation}
\int\langle T_i^i \rangle dV= 0, \quad E = -\int\langle T_\mu^\mu \rangle dV,
\label{QFT-vir}
\end{equation}
where the expectations values are taken with respect to a normalized
stationary state representing a bound state in its rest frame.  

If a semiclassical (perturbative) expansion is meaningful,
Eqs.~(\ref{QFT-vir}) amount to the classical field theory virial theorem,
namely, Eq.~(\ref{Tii}) or Eq.~(\ref{E=T}), plus quantum corrections.  The
simplest quantum correction consists in employing the limited Fock space
approximation \cite{FHM}, which is a variational approximation equivalent to
the Schr\"odinger equation in a limited Fock space that does not include
renormalization effects.  When applied to an electron-positron bound state
\cite{DP}, the result is that its energy $E$ can be expressed by either
Eq.~(\ref{L-S}), without $U$ terms and with $m_1=m_2$, of course, or by
Eq.~(\ref{E-int}). In both equations, the classical motion is replaced with a
probability distribution, given by a quantum wave-function for the former and
by a classical ``mass density'' for the latter.  The limited Fock space
approximation for bound states is connected with the standard treatment of
bound states in QFT, which involves several approximations that lead to the
Bethe-Salpeter equation \cite[\S\hspace{2pt}6]{Greiner}.

In the limited Fock space approximation for spinor QED no infinities arise
\cite{FHM,DP}, but, in general, one must consider the infinities that
inevitably arise in QFT.  Those infinities require regularization with an
ultraviolet (UV) cutoff, which necessarily breaks scale invariance.  However,
let us remark that some infinities already arise in a classical field theory
with point-like particles and because of this problem the classical
electrodynamics of particles of mass $m$ and charge $e$ is not consistent on
scales smaller than the classical radius $e^2/(mc^2)$ \cite{LL2,Jack}.  In
fact, as explained in Sec.~\ref{em-virial}, the classical relativistic virial
theorem holds only after subtraction of the infinite self-energy of point-like
charges.  Since the classical radius is smaller than the Compton wavelength
$\hbar/(mc)$, where the quantum effects take over, the regularization of
infinities is essentially a quantum problem.%
\footnote{This statement must be qualified: In the exceptional case in which
  the strong interaction holds a charge $Ze$ in a point-like nucleus, $e^2$
  has to be replaced with $Ze^2$, and the condition $Z\a> 1$ precisely means
  that $Ze^2/(mc^2)$ is larger than $\hbar/(mc)$.}  For the virial theorem in
the form of Eq.~(\ref{QFT-vir}), the relevant quantum effects of
renormalization manifest themselves in the energy-momentum tensor trace
anomaly, as pointed out by Dudas and Pirjol \cite{DP}.

\subsection{The QED virial theorem and the trace anomaly}

Before considering the renormalization process for bound states in QFT, let us
recall why renormalization is necessary for the classical virial theorem to
hold (Sec.~\ref{em-virial}).  For a set of electromagnetically bound
particles, fulfilling Eq.~(\ref{Tii-N}), the virial theorem of Eq.~(\ref{Tii})
implies that their positive kinetic ``pressure'' must be balanced by negative
electromagnetic stresses.  Therefore, their electrostatic energy, as given by
Eq.~(\ref{Coul}), must be negative, despite that it appears to be positive.
In fact, this energy is actually infinite, but it can become negative after
suitable subtractions, absorbed by {\em mass renormalizations}.  Mass
renormalization is also necessary in QFT because of the appearance of
infinite self-energies, but the divergence structure is substantially
modified. Besides, in QFT, the charge must be renormalized as well.

In a semiclassical expansion, the calculation of the first quantum correction
to some bound state energy only requires the calculation of the small
oscillations of the corresponding classical solution.  For example, for a
heavy atom, this semiclassical approach is equivalent to the old Bloch
hydrodynamic treatment of the Thomas-Fermi atom model \cite{BWF}.  However,
this model is non-relativistic.  Proper relativistic examples are provided by
classical relativistic field theory localized solutions (``classical lumps'')
\cite[Ch~6]{Coleman}.  In general, the calculation of quantum corrections
begins with the calculation of the stability matrix determining the small
oscillations.  The total energy is the classical energy plus the contribution
of these small oscillations.
Once these oscillations are quantized, the first quantum correction to 
the classical energy is
$$
\d E = \frac{\hbar}{2} \sum_i  \o_i\,, 
$$
where the sum goes over the oscillation modes and $\o_i$ is the angular
frequency of the $i$-th mode.  This sum diverges at high frequency (in the
UV).  The modes can be labeled by three independent numbers that can be
assembled, for large values, in a wave-number vector $\bm{k}$. Therefore, the
UV divergence can be segregated by expressing the sum as an integral over
$\bm{k}$, for large $k$, and taking $\o_i \approx c k$, as corresponds to free
modes; namely,
$$ 
\d E = \hbar c V \int^{\L} \left[k + \Mfunction{O}(1)\right]
\frac{d^3k}{(2\pi)^3} 
= \frac{\hbar c}{8\pi^2} V \left[\L^4 + \Mfunction{O}(\L^3)\right] ,
$$ 
where $V$ is the system's volume, a UV cutoff $\L$ has been introduced, and a
sum over two polarizations has been made, assuming that the free modes
correspond to photons.  One could make $\d E$ finite and attribute it a
physical meaning by choosing a physical cutoff $\L$, such as $\L \sim
mc/\hbar$, namely, the inverse of the electron Compton wavelength.  At any
rate, the leading term, which is positive and proportional to $\L^4$, is
present in the absence of matter and must be subtracted.  The subleading
divergent terms depend on the detailed spectrum $\o_i\,$ and hence on the
matter state, and they can have either sign. After renormalization (or
definition of a finite $\L$), these terms give rise to measurable
electromagnetic energies.

The divergent parts of the vacuum energy indeed have to be subtracted in the
calculations of electromagnetic energies, for example, in the calculation of
Casimir or Van der Waals forces \cite{Milonni}.  For illustration, let us
briefly consider a case in which the frequencies $\o_i$ are easy to calculate,
namely, the case of a dilute gas of $N$ atoms per unit volume in a box of
volume $V$ \cite[\S\hspace{2pt}3.7]{Milonni}. The allowed frequencies are the
free-field frequencies modified by the refractive index of the gas, $n(\o)$,
i.e. $\o_i \approx ck/n$.
If we assume, for simplicity, that there is only one resonant frequency,
$\o_0\,,$ we can take
$$
n(\o)= 1 + \frac{N e^2}{2 m (\o_0^2 - \o^2)}\,.
$$ 
Therefore,
$$
\d E = \frac{\hbar c V}{(2\pi)^3} \int \frac{k\,d^3k}{n(k)} = 
\frac{\hbar c V}{2\pi^2} 
\left[ \frac{\L^4}{4} + \frac{N e^2\L^2}{4 m c^2}  + 
\frac{N e^2}{4 m c^2} \left( k_0^2 + \frac{N e^2}{2 m c^2} \right) 
\ln\left|\frac{\L^2}{k_0^2 + N e^2/(2 m c^2)} - 1\right| \right].
$$
The quadratic divergence is proportional to the number of atoms, $NV$, but is
independent of $\o_0$, and the proportionality constant only depends on
fundamental constants. Indeed, this term corresponds to the energy of $NV$
{\em free} electrons, and, if we take again $\L \sim mc/\hbar$, the energy per
electron is of the order of $e^2/(\hbar/mc)$, that is to say, of the electron
quantum self-energy.  Therefore, this divergence can be absorbed by mass
renormalization.  The logarithmic term is of smaller magnitude and depends on
$\o_0$.  If $N e^2/m \ll \o_0^2$ (the dilution condition), the logarithmic
term can be identified with the Lamb shift \cite[\S\hspace{2pt}3.7]{Milonni}.
Notice that the quantum corrections are, after renormalization, not only
finite but small (of order $\a$).  If we take, instead of $\L \sim mc/\hbar$,
the much smaller cutoff $\L = k_0$, so that $n(k) > 1$ over the full range of
integration, then $\d E$ is negative, as corresponds to attractive Van der
Waals forces.

The result of the preceding calculation of $\d E$ and, in particular, its
cutoff dependence are typical of {\em one-loop effective potential}
calculations.  Indeed, the energy $E$ can be generally calculated in terms of
the minimum of the effective action corresponding to the bound state.  The
$\L^4$ term is already present in the vacuum, when a generic regularization
method is employed, and leads to the well-known {\em cosmological constant}
problem.  In this regard, Ossola and Sirlin \cite{Os-Sir} study the
contribution of fundamental particles to the vacuum energy density, comparing
various regularization methods, and conclude that, for non-interacting
particles, the divergence can be made quadratic rather than quartic (no $\L^4$
term), and that massless particles do not contribute. This result follows from
properly considering relativistic covariance and scale invariance of free
field theories in the massless limit and, therefore, calculating $\langle
T^{00} \rangle$ or any other component of $\langle T^{\mu\nu}\rangle$ in terms
of the trace $\langle T^{\l}_\l \rangle$, by employing the relation $\langle
T^{\mu\nu} \rangle= \langle T^{\l}_\l \rangle g^{\mu\nu}/4.$ The divergences
of the trace $\langle T^{\l}_\l \rangle$ are easily obtained for free fields,
but, with interactions, new divergences appear, namely, new quartic, quadratic
and logarithmic divergences.  After renormalization, the energy-momentum
tensor trace has quantum corrections, that is, a \emph{trace anomaly} appears.
In particular, the trace is not zero even in the massless case.  As the virial
theorem can be expressed in terms of the energy-momentum tensor trace, one
concludes that the virial theorem must include a quantum trace anomaly term
\cite{DP}, since this anomaly appears even in the vacuum.

The general form of the fermionic QED trace anomaly was computed by Adler et
al \cite{Adler}:  
\begin{equation}
T_{\mu}^{\mu}= -K_1 m_0 c^2\bar{\psi}\psi + K_2 N[F_{\l\s}F^{\l\s}],
\label{T-QED}
\end{equation}
where $N$ indicates a type of normal product (see \cite{Adler}) and
\begin{eqnarray*}
K_1 = 1 + \d(\a) = 1 + \frac{3\a}{2\pi} + \cdots,\\
K_2 = \frac{1}{4} \b(\a) = \frac{1}{4}\left(\frac{2\a}{3\pi} +
\frac{\a^2}{2\pi^2} + \cdots\right).
\end{eqnarray*}
The functions $\d$ and $\b$ are associated to the anomalous dimension of the
fermion field $\psi$ and to the coupling constant renormalization,
respectively.  Notice that the first term of the trace anomaly consists of the
classical part in Eq.~(\ref{T-fermi}) plus quantum corrections due to mass
renormalization, whereas the second term is purely of quantum origin and is
due to charge renormalization.

The QED trace anomaly is indeed small, due to the smallness of $\a$, which
makes perturbation theory work.  This implies that the renormalized value of
the quantum correction $\d E$ to a classical bound-state energy is small as
well.  Fundamentally, it implies that the vacuum is not altered, that is to
say, that there is no vacuum decay.  Naturally, the vacuum decay in the strong
field of a nucleus is an exception, due to the actual coupling constant,
$Z\a$, not being small (Sec.~\ref{ultra}). At any rate, it is interesting to
consider general strongly-bound states. A state can be considered strongly
bound if the largest part of its energy is due to the interaction of its
constituents rather than to their rest masses. This condition is naturally met
by hadronic states, namely, the states of quarks bound by the strong
interaction described by QCD.

\subsection{Scale symmetry breaking, Callan-Symanzik equations, and QCD}
\label{QCD}

QCD is the theory of the strong interaction, although effective low-energy
theories, such as the meson-nucleon interaction theory, are still useful.  The
meson-nucleon theory that only includes the lightest mesons is somewhat
similar to the photon-electron theory in QED, except for the finite range of
the mesons.  However, the magnitude of the meson-nucleon coupling is
$g^2/(4\pi\hbar c) \simeq 1$, which should imply that the bound states are
ultrarelativistic and their energies nearly vanish! This does not happen, of
course, because scale invariance is badly broken by quantum effects.
Coleman \cite[Ch~3]{Coleman} employs the meson-nucleon model for illustrating
the breaking of scale invariance, by analyzing the perturbative behavior of
correlation functions in the deep Euclidean region: in addition to simple
powers of the scale (given by the field dimensions), there appear logarithms
as well. Then, Coleman shows that some series of logarithms can be absorbed
into anomalous field dimensions, while other logarithms remain, but the series
of the latter can be absorbed by a renormalization of the couplings.
In general, the scale dependence introduced by the renormalization process can
be expressed in terms of a set of simple differential equations for the
correlation functions, namely, the renormalizations group equations, while the
effect of scale-symmetry breaking is expressed by the Callan-Symanzik
equations. Naturally, both sets of equations are related \cite[Ch~3]{Coleman}.
There is an infinite set of Callan-Symanzik equations, one equation for every
correlation function.  To apply these equations to bound states, it is useful
to realize that they all can be derived from a master Callan-Symanzik
equation, which is, in addition, connected with the energy-momentum tensor
trace anomaly.

To obtain the master Callan-Symanzik equation, let us introduce the 
generating functional 
$$Z[\l^j,g_{\mu\nu}]= \int \D\f\, \exp(iS[\f,\l^j])\,,$$
namely, the vacuum transition amplitude, 
where the action $S$ depends on a set of fields $\f$ and coupling constants
$\l^j$ and is defined in a curved space-time, and let us also introduce the
functional $W = -i \log Z$.  The effect of a scale transformation can be
alternately realized through the metric, so
$$l\frac{dW}{dl} = 2\int d^4x\, g^{\mu\nu}(x)\frac{\d W}{\d g^{\mu\nu}(x)} 
= \int d^4x \sqrt{\det g_{\mu\nu}} \,\langle T_{\mu}^{\mu} \rangle,$$
where $l$ is the scale, or through the coupling constants, so
$$l\frac{dW}{dl} = \sum_i\b^i(\l) \int d^4x \sqrt{\det g_{\mu\nu}}\, \langle
\O_i \rangle + \A,$$ 
where $\b^i(\l) = l \,d\l^i/dl$ (``beta functions''), $\{\O_i\}$ is the set of
interaction terms (a subset of ``composite fields''), fulfilling
$$\frac{\d W}{\d\l^i} =\int d^4x \sqrt{\det g_{\mu\nu}}\,\langle \O_i
\rangle,$$ 
and $\A$ is the extra anomalous term that arises in a curved geometry.
Therefore, in flat space-time,
\begin{equation}
  \int d^4x \,\langle T_{\mu}^{\mu} \rangle
  = \sum_i\b^i(\l) \int d^4x  \,\langle \O_i \rangle .
\label{CS}
\end{equation}
This equation%
\footnote{This equation is well known. It plays an important role in 2d field
  theory \cite{Zamo1,Zamo2} and it is discussed, in general, by Osborn
  \cite{Osborn}.
}
generates an infinite hierarchy of equations for the correlations of the
$\O_i$ by taking derivatives of it with respect to the couplings
$\l^i$. Equations for the correlations of fields other than $\O_i$ and, in
particular, of the elementary fields $\f$ can also be obtained by introducing
in $W$ the respective sources.  These equations are in essence the standard
Callan-Symanzik equations.  The master Callan-Symanzik equation (\ref{CS})
shows the general form of the trace anomaly \cite{Osborn}.  For example, in
QED, Eq.~(\ref{CS}) is equivalent to the operator equation (\ref{T-QED}). To
see this, one has to notice that the set $\{\l^i\}$ of coupling constants
includes dimensionful constants.  However, the dimensionless coupling
constants play a special role, because they can contribute to each $\b^i(\l)$,
whereas the dimensionful coupling constants (corresponding to
super-renormalizable interactions, e.g., mass terms) can only contribute to
some of them, as a simple power-counting argument shows. Indeed, in
Eq.~(\ref{T-QED}), $\a$ contributes to both terms but $m_0$ is only allowed in
the first term and with exponent one. Also, notice that the ``beta functions''
of dimensionful coupling constants are classically non-vanishing and give rise
to the classical value of $T_{\mu}^{\mu}$, whereas those of dimensionless
coupling constants are purely ``anomalous.''

Equation (\ref{CS}) can be generalized by replacing the vacuum expectation
values with the expectation values in the stationary state corresponding to a
bound state, namely, by introducing the appropriate boundary conditions in the
generating functional $Z$.  For stationary states, the four-dimensional
integrals in Eq.~(\ref{CS}) become spatial integrals and the equation directly
yields the quantum corrections to the relativistic virial theorem (\ref{E=T}),
in terms of the beta functions and the expectation values $\langle \O_i
\rangle$.  The beta functions can be calculated in perturbation theory, but
the $\langle \O_i \rangle$ are harder to calculate, because they are
essentially non-perturbative.

To appreciate the importance of quantum corrections to the relativistic virial
theorem, let us consider the theory of hadrons in QCD.  As is well known, the
crucial difference between QED and QCD is that their beta functions have
opposite signs, so that QCD is {\em asymptotically free}, namely, the
interaction vanishes at high momenta.  In contrast, quarks interact strongly
at low momenta, and, in fact, hadrons made of light quarks have a size such
that the contributions of the interactions to their energy are much larger
than the contribution of the quark masses. At the same time, the contribution
of the trace anomaly is crucial.  To clarify all this, let us consider a
concrete hadron model.

In the MIT bag model \cite{bag}, a constant positive potential energy $B$, per
unit volume, is added to a free-field Lagrangian density inside a finite part
of space.  A semiclassical description of this simple field theory constitutes
a sort of statistical model of hadrons.  In the MIT bag model, the particles,
named then ``partons'' but now identified with quarks and gluons, are free and
massless inside the bag and they move with the speed of light. Therefore,
hadrons are truly ultrarelativistic bound states, as studied in
Sec.~\ref{ultra}.  The vacuum inside the bag takes place for small distances
and high momentum and hence corresponds to the QCD perturbative vacuum.  As
the particles inside the bag separate from one another, they enter the strong
coupling phase, in which the perturbative vacuum is not stable and decays to
the standard QCD vacuum (a condensate of quark-antiquark pairs).  Vice versa,
the QCD vacuum becomes unstable at high momenta and decays to the perturbative
vacuum.  The transition is assumed to take place over a very small distance,
at the bag's surface.  In the statistical model, the confining interaction is
just modeled as a vacuum pressure, namely, the difference between the null
vacuum pressure outside the bag and the {\em negative} ``vacuum pressure'' in
the bag. The effect of this pressure difference is just to keep the particles
inside the bag, as if the difference of pressures were produced by the bag
surface.

The application of the virial theorem to the gas of free massless particles of
the statistical bag model is straightforward: the theorem just states that the
energy density is three times the pressure difference
\cite[\S\hspace{2pt}35]{LL2}.  On the other hand, the energy-momentum tensor
of the perturbative vacuum is constant and proportional to the metric
$g^{\mu\nu}$, as corresponds to a constant term in the Lagrangian.  Therefore,
we have the following relations: the virial relation
$E_\mathrm{quarks+gluons}/V=3P_\mathrm{quarks+gluons}$ (where $V$ denotes the
bag's volume), the balance of pressure in the bag,
$P_\mathrm{quarks+gluons}=-P_\mathrm{vac}$, and the vacuum relation
$E_\mathrm{vac}=-P_\mathrm{vac}V$. Let us remark that the balance of pressure 
is also a virial relation, namely, a particular case of 
$\int \overline{T^{i}_i}\, dV= 0$.  
All the preceding equations together imply that
\begin{equation}
E_\mathrm{quarks+gluons}= 3E_\mathrm{vac}\,, \quad E=E_\mathrm{quarks+gluons}+
E_\mathrm{vac} = 4E_\mathrm{vac} = 4BV.
\label{bag-vir-rel}
\end{equation} 
Notice that the confining
interaction accounts for one fourth of the total energy, like in the classical
model of a particle with electric charge of Sec.~\ref{em-virial} that has a
confining Poincar\'e energy-momentum tensor proportional to $g^{\mu\nu}$.

The bag surface has been taken to be fixed but this is not necessary: the only
change when the bag surface is allowed to move is that $V$ must be replaced
with the average $\overline{V}$, so $E = 4 B \overline{V}$. This equation is
also obtained by Chodos et al \cite{bag} as a relativistic virial theorem
specially tailored for the bag model Lagrangian and proved without employing
the energy-momentum tensor.

Since the (improved) classical energy-momentum tensor is traceless in the MIT
bag model, confinement is due to the quantum energy-momentum tensor trace part
(the part proportional to $g^{\mu\nu}$), namely, to the the trace anomaly.  In
the MIT bag model, this quantity is spatially constant, but in QCD the trace
anomaly is given by an equation similar to Eq.~(\ref{T-QED}) that includes
suitable color indices; namely, it contains the term $\bar{\psi}m\psi$, where
$m$ is the quark mass matrix, and the term $F^2$, with the color indices
contracted.  According to the virial theorem (\ref{QFT-vir}) and assuming that
the quarks are massless, the energy is given by
the spatial integral of $-\langle T^{\l}_\l \rangle = -\b(g) \langle
F^2\rangle/(2g)$  (in the remainder, we take $\hbar= c= 1$). 
Therefore, at first order in $\a_\mathrm{s} = g^2/(4\pi)$,
$$E = -\int dV \langle T^{\l}_\l \rangle = 
\frac{9\a_\mathrm{s}}{8\pi} \int dV\langle F^2 \rangle$$ (for three quark
flavors).  The last equation is a particular case of Eq.~(\ref{CS}), for a
stationary state and only one coupling.  The required spatial integral,
$$\int \langle F^2 \rangle dV = -2 \int (\bm{E}^2 - \bm{B}^2) dV,$$
must be calculated with the bag boundary conditions.  For the lowest energy
states, the integral can be calculated in terms of the respective integrals of
$\bm{E}^2$ and $\bm{B}^2$, which also appear in the calculation of the gluonic
corrections to the ground states of the MIT bag model
\cite[\S\hspace{2pt}3.12]{GreinerSS}.  These integrals contain divergent terms
that correspond to self-interactions that are absorbed by mass
renormalizations.  The final result is that the electric integral vanishes and
the magnetic integral is proportional to $1/R$, namely, to the inverse of the
spherical bag radius, with a coefficient of proportionality of the order of
unity, as one should expect.  Indeed, the partons in the lowest excitation
states of the bag must have energy of the order of $NV^{-1/3}$, where $N$ is a
small integer, as follows from dimensional analysis \cite{bag}. The trace
anomaly calculation cannot specify $N$, since $\a_\mathrm{s}$ is undetermined.

Taking any one of the lowest energy hadronic states, the bag relation 
$E= 4BV \sim {N}/{R},$ 
with $V \sim R^3$, implies that $B \sim R^{-4}$, where $R$ is the hadron
radius.  Therefore, the fundamental dimensional parameter $B$ can be traded
for $R$.  However, let us introduce instead a fundamental QCD dimensional
parameter, namely, the RG invariant $\L_\mathrm{QCD}$, which naturally arises
in massless QCD through the renormalization process, constituting what is
called {\em dimensional transmutation}.  $\L_\mathrm{QCD}$ can be defined, for
example, as the scale at which the running QCD coupling equals unity,
$\a_\mathrm{s}(\L_\mathrm{QCD}) = 1$.  The scale $\L_\mathrm{QCD}$ determines
the size of the lowest excitation states of hadrons, namely, $R \sim
1/\L_\mathrm{QCD}$.  Therefore, we have $B \sim \L_\mathrm{QCD}^{4}$ and
$E_\mathrm{vac} = BV \sim \L_\mathrm{QCD}$. Also, $E \sim \L_\mathrm{QCD}$.
Naturally, $R$ and $E$ can be expressed in terms of $\L_\mathrm{QCD}$ for
general hadron models with massless quarks, not only for the MIT bag model, as
deduced from dimensional transmutation.  Dimensional transmutation is a
consequence of the breaking of scale invariance, which also gives rise to the
trace anomaly.
However, let us remark that the virial theorem only allows us to establish
{\em one} relation, which is {\em exact} and says, in essence, that the
confining term, namely, the trace anomaly, accounts for one fourth of the
total energy, for massless quarks, as in Eq.~(\ref{bag-vir-rel}).  This exact
relation derives from the obvious equation
$T^{\mu\nu}=T^{\mu\nu}_{\mathrm{quarks+gluons}} + T_\a^\a g^{\mu\nu}/4$,
where, classically, $T_\a^\a=0$, and
$$
T^{\mu\nu}_{\mathrm{quarks+gluons}} = \frac{i}{2} \left(
\bar{\psi} \g^{(\mu}\D^{\nu)} \psi -\D^{*(\mu}\bar{\psi} \g^{\nu)} \psi 
\right) - \frac{g^{\mu\nu}}{4} F^2 + F^{\mu\s}\cdot{F^\nu}_{\s}\,.
$$
Therefore, on account of the QFT virial theorem, Eq.~(\ref{QFT-vir}),
\begin{equation}
E = 
\int \langle T^{00}_{\mathrm{quarks+gluons}} \rangle\,dV + \frac{E}{4}\,,
\label{virialQCD}
\end{equation}
in analogy with Eq.~(\ref{virialE}).  Naturally, $T_\a^\a=0$ implies $E=0$,
but the trace anomaly makes $E$ non-vanishing.  Equation (\ref{virialQCD}) is
exact for massless QCD and is approximately valid for hadrons made of
light quarks.  On the other hand, there is another relation, always
approximate, that determines the size of the lowest excitation states of
hadrons in terms of a fundamental parameter, say, $\L_\mathrm{QCD}$ (or $B$,
in the bag model).  Obviously, the latter relation cannot hold for high
excitation states, whereas the virial theorem does.

In more elaborate hadron models, one can consider quark masses,
$m_\mathrm{Q}$, so that the classical energy-momentum tensor is not
traceless. Nevertheless, for light quarks, namely, the {\em u} and {\em d}
quarks (and surely the {\em s} quark as well), $m_\mathrm{Q} \ll
\L_\mathrm{QCD}$, so the trace anomaly is still dominated by the $F^2$
term. If Nature were such that $\L_\mathrm{QCD} \ll m_\mathrm{Q}$ for all
quarks, then QCD, in spite of still having the property of quark confinement
and still being asymptotically free, would share some features of QED: all
hadrons would actually be non-relativistic quark bound states.

\section{Conclusions}

The virial theorem expresses the condition of average equilibrium of a bound
state, namely, the condition that its average shape stays constant and, in
particular, that its average size stays constant. The former condition is
expressed as the tensor virial theorem, whereas the latter is expressed as the
ordinary scalar virial theorem, Eq.~(\ref{Tii}), which is actually the trace
of the tensor virial theorem. This theorem implies that the positive pressure
of particles or fields corresponding to a traceless energy-momentum tensor
must be compensated for by the negative stresses corresponding to the trace
part of the energy-momentum tensor.  Eq.~(\ref{Tii}) is valid in both
classical and relativistic physics, but in the latter the equivalent form in
terms of the energy-momentum tensor trace, Eq.~(\ref{E=T}), is more
convenient. This equation implies that the proportion of the bound state
energy corresponding to a traceless energy-momentum tensor is three
fourths. This fraction is connected, in particular, with the $4/3$ that
appears in the classic problem of electromagnetic inertia.

Both the trace of the stress tensor and the trace of the energy-momentum
tensor are related to the generators of, respectively, space and space-time
dilatations.  Naturally, the connection with dilatations arises because these
are the transformations that change the size of the system. The virial theorem
shows that the average size of the system is determined by its energy $E$.
Full space-time scale invariance only takes place when the (average of the)
trace of the energy-momentum tensor vanishes, which corresponds to $E=0$ and,
in principle, to the vacuum.

While non-relativistic mechanics admits similarity of motion for kinetic and
potential energies that are homogeneous functions of their respective
variables, this symmetry is lost in relativistic mechanics, in which the
kinetic energy is never a homogeneous function.  However, there exists a
relativistic notion of similarity, which is such that velocities are left
invariant but masses scale as energies do.  Therefore, the only situation in
which full scale invariance can appear is in the ultrarelativistic domain,
when the masses are set to zero and the virial theorem implies that the energy
vanishes.  This ultrarelativistic scale invariance appears, for example, in
the strong electric field of a heavy nucleus.

On the other hand, only in the context of quantum field theory does scale
invariance acquire its deepest meaning, since there is naturally only one
scale, either length or mass. In a theory with massless fundamental particles,
bound systems should have $E=0$ and be massless as well, so there should be no
scales and the theory would naturally be scale invariant. However, in quantum
field theory the vacuum is non trivial, and there arise virtual
particle-antiparticle pairs that make quantum contributions to energies that
are actually infinite and, after renormalization, bring about a scale
dependence, expressed by the Callan-Symanzik (or renormalization group)
equations. This dilatation symmetry breaking can also be expressed in terms of
the energy-momentum trace anomaly, which must be included in the quantum field
theory virial theorem. Therefore, this theorem is just a generalization of the
Callan-Symanzik or trace anomaly equations that is applicable to bound
states. Full scale invariance takes place only at renormalization group fixed
points, where the virial theorem becomes trivial.

The QED energy-momentum tensor trace anomaly is proportional to (powers of)
the coupling constant $\a=1/137$, so it is small. In general, the quantum
corrections to relativistic bound states are small in QED. As those bound
states are weakly coupled, they are weakly relativistic as well.
Exceptionally, the strong nuclear interaction within a heavy nucleus creates a
concentration of positive charges such that their electromagnetic field is
sufficiently strong to produce important relativistic effects, in particular,
a qualitative change in the quantum vacuum around the nucleus: an electron
bound to the nucleus can have {\em negative} energy, and the bound state
actually decays to a new vacuum state with zero energy.

If the QED coupling $\a$ were sufficiently large, then vacuum instability
would not be exceptional, because then positronium could have a negative
energy state, and the standard QED vacuum would be unstable against production
of electron-positron pairs. A phenomenon of this kind takes place in QCD, so
the QCD vacuum is not the perturbative vacuum but a condensate of
quark-antiquark pairs. This phenomenon is associated with the magnitudes of
dilatation symmetry breaking and of the trace anomaly in QCD.  In fact, the
existence of hadrons in massless QCD is conditioned by the trace anomaly,
which is the essential ingredient of the virial theorem in this case.  The
dilatation symmetry breaking can be expressed as a dimensional transmutation,
which gives rise to a scale, the renormalization group invariant
$\L_\mathrm{QCD}$. This scale is definitely larger than the light quark masses, 
which implies that
the essential properties of the hadrons formed by them are described by
massless QCD.  Assuming that quarks are massless, the virial theorem implies
that one fourth of the energy (or mass) of a hadron comes from the trace
anomaly term whereas three fourths come from quarks and gluons.  The bag model
of hadrons is useful for a straightforward application of the virial theorem
and the computation of the trace anomaly, as has been shown.

A different type of strong interaction occurs in astrophysics, namely, in
compact bodies, where the additive nature of gravity leads to strong
gravitational fields, besides strong quantum effects.  However, the
generalization of the virial theorem to General Relativity presents problems
that are beyond the scope of this paper, as cautioned in the
introduction. Nevertheless, we can affirm that the master Callan-Symanzik
equation (\ref{CS}) that includes the curvature dependent trace anomaly $\A$
must play an important role in such a generalization.

\subsection*{Acknowledgments}

I thank Sergey Apenko and Andrey Semenov for comments on the manuscript.


\begin{thebibliography}{99}

\bibitem{Collins} Collins G W 1978 \emph{The Virial Theorem in Stellar
    Astrophysics} 
(Tucson, AZ: Pachart)

\bibitem{LL1}
Landau L and Lifshitz E 1976 \emph{Mechanics} 3rd edn (Oxford: Pergamon) 

\bibitem{Low}
L\"owdin P O 1959 
\emph{J. Mol. Spectrosc.} 3 46–-66

\bibitem{AB}
Andersen C M and von Baeyer H C 1971 
\emph{Am. J. Phys.} 39 914–-9

\bibitem{Kleban}
Kleban P 1979 
\emph{Am. J. Phys.} 47 883-–6

\bibitem{Kampen}
van Kampen N G 1972 
\emph{Rep. Math. Phys.} 3 235–-9 

\bibitem{Nach}
Nachtergaele B and Verbeure A 1986 
\emph{J. Geom. Phys.} 3 315–-25

\bibitem{Blud-Ke}
Bludman S and Kennedy D C 2011 \emph{J. Math. Phys.} 52 042902

\bibitem{LL2}
Landau L and Lifshitz E 1962 \emph{The Classical Theory of Fields} 2nd edn
(Oxford: Pergamon) 

\bibitem{Raf}
Rafelsky J 1977 \emph{Phys. Rev.} D 16 1890

\bibitem{Brack}
Brack M 1983 \emph{Phys. Rev.} D 27 1950

\bibitem{DP}
Dudas E A and Pirjol D 1991 \emph{Physics Letters} B 260 186

\bibitem{Gold}
Goldstein H 1950 \emph{Classical Mechanics} (Cambridge Mass: Addison-Wesley) 
p~214

\bibitem{Jack} Jackson J D 1999 {\em Classical Electrodynamics} 3rd
  edition (John Wiley)

\bibitem{a-d} Anderson J L 1967 \emph{Principles of Relativity Physics} (NY:
  Academic Press)

\bibitem{Barut} Barut A O 1980 \emph{Electrodynamics and Classical Theory of
    Fields and Particles} (NY: Dover)

\bibitem{Aus}
Lucha W and Sch\"oberl F F 1990 \emph{Phys. Rev. Lett.} 64 2733

\bibitem{Kor}
Hwang D S, Kim C S and Namgung W 1997 \emph{Phys. Lett.} B 406 117

\bibitem{Laue}
von Laue M 1911 \emph{Annalen Phys.} 35 524

\bibitem{Ohanian}
Ohanian H C 2009 \emph{Studies in History and Philosophy of Modern Physics} 40
167--173 

\bibitem{Hawk-Ell}
Hawking S W and Ellis G~F~R 1973
\emph{The large scale structure of space-time} (Cambridge U. P.)

\bibitem{BB} 
Bialynicki-Birula I 1983 
\emph{Phys. Rev.} D 28 2114 

\bibitem{Fock}
Fock V 1930 
\emph{Zeitschrift f\"ur Physik} 63 855--858

\bibitem{RW} 
Rose M E and Welton T A 1952	
\emph{Phys. Rev.} 86 432--433

\bibitem{CED}
Wakano M 1966 \emph{Prog. Theor. Phys.} 35 1117--1141 

Radford C 
2003 \emph{J. Phys. A} 36 5663--5681

\bibitem{W} Weinberg S  1995 \emph{The Quantum Theory of Fields} vol~I
  (Cambridge U. P.)  

\bibitem{Hobart}
Hobart R H 1963 \emph{Proc.\ Phys.\ Soc.}~82 201--203

\bibitem{Derrick}
Derrick G H 1964 \emph{J.\ Math.\ Phys.}~5 1252--1254

\bibitem{Coleman}
Coleman S 1985 \emph{Aspects of Symmetry} (Cambridge U. P.)

\bibitem{CCJ}
Callan C G Jr, Coleman S and Jackiw R 1970 
	\emph{Annals of Physics} 59 42--73


\bibitem{Herb}
Herbst I W 1977 \emph{Commun.\ Math.\ Phys.}~53 285--294;
\emph{Commun.\ Math.\ Phys.}~55 316 (addendum)

\bibitem{Greiner}
Greiner W and Reinhardt J 2003 \emph{Quantum Electrodynamics}
(Berlin: Springer-Verlag)

\bibitem{FHM}
Finger J, Horn D and Mandula J E 1979
\emph{Phys. Rev.} D 12 3253

\bibitem{BWF}
Ball J A, Wheeler J A and Firemen E L 1973 
\emph{Rev. Mod. Phys.} 45 333

\bibitem{Milonni}
Milonni P W 1994 \emph{The Quantum Vacuum: an introduction to Quantum
Electrodynamics} (San Diego, CA: Academic Press)

\bibitem{Os-Sir}
Ossola G and Sirlin A 2003 \emph{Eur. Phys. J.} C 31 165

\bibitem{Adler}
Adler S L, Collins J C and Duncan A 1977 \emph{Phys. Rev.} D 15 1712

\bibitem{Zamo1}
Zamolodchikov A B 1986 \emph{JETP Lett.} 43 730--732

\bibitem{Zamo2}
Zamolodchikov A B 1987 \emph{Sov. J. Nucl. Phys.} 46 1090--1096

\bibitem{Osborn}
Osborn H 1991 \emph{Nucl.\ Phys.} B 363 486--526


\bibitem{bag}
Chodos A et al 1974 \emph{Phys. Rev.} D 9 3471

\bibitem{GreinerSS}
Greiner W, Schramm S and Stein E 2002 \emph{Quantum Chromodynamics}
(Berlin: Springer-Verlag)

\end{thebibliography}
\end{document}